\documentclass[12pt]{article}
\usepackage[english]{babel}
\usepackage[utf8]{inputenc}
\usepackage{latexsym}
\usepackage{dsfont}
\usepackage{amsfonts,amsbsy,bm,euscript,mathrsfs}
\usepackage{amssymb,stmaryrd,faktor}
\usepackage[tbtags]{amsmath}
\usepackage{amsthm}
\usepackage[nosort]{cite}
\usepackage{setspace}
\usepackage[toc,page]{appendix}

\usepackage[hidelinks]{hyperref}
\hypersetup{
colorlinks=false,
citecolor= blue,
linkcolor= blue,
urlcolor= blue,
breaklinks=true
}

\allowdisplaybreaks

\numberwithin{equation}{section}

\begin{document}


\begin{titlepage}
\begin{center}
\vspace*{-1.0cm}
\hfill {\footnotesize ZMP-HH/23-15}

\vspace{2.0cm}

{\LARGE  {\fontfamily{lmodern}\selectfont \bf Jordan blocks and the Bethe ansatz III: \\ \vspace{0.25cm} Class 5 model and its symmetries}} \\[.2cm]

\vskip 1.5cm
\textsc{Juan Miguel Nieto Garc\'ia\footnote{\href{mailto:juan.miguel.nieto.garcia@desy.de}{\texttt{juan.miguel.nieto.garcia@desy.de}}}}
\\
\vskip 1.2cm

\begin{small}
\textit{$^1$II. Institut für Theoretische Physik, Universität Hamburg\\
Luruper Chaussee 149, 22761 Hamburg, Germany.}
\end{small}

\end{center}

\vskip 0.7 cm
\begin{abstract}
\vskip0.5cm

\noindent We study the Hilbert space of the Class 5 model described in \href{https://arxiv.org/abs/1904.12005}{arXiv:1904.12005}. Despite being integrable, neither its transfer matrix nor its Hamiltonian are diagonalisable, meaning that the usual Algebraic Bethe Ansatz does not provide the full Hilbert space. Instead, we make use of the symmetries of the model to construct the Jordan blocks of the transfer matrix. We also show that the Hamiltonian and the transfer matrix, despite commuting, do not have the same Jordan block structure.

\end{abstract}

\end{titlepage}


\section{Introduction}

Non-Hermitian systems are used to model many physical phenomena, from electric circuits with diodes to the absorption of electromagnetic waves propagating through a material (see \cite{Ashida:2020dkc} and references therein for more examples). Despite that, classical non-Hermitian models are less studied than Hermitian models due to their complexity. This is even more prominent in quantum mechanics, where the Dirac–von Neumann axioms specifically demand observables to be self-adjoint. However, self-adjointness is actually not a necessary condition for a Hamiltonian to have a real energy spectrum, and it can be relaxed to demanding the existence of an anti-linear operator that commutes with it. The most common choice in the literature is the $\mathcal{PT}$ symmetry operator \cite{Bender:1998ke,Bender:2007nj,Alexandre:2015kra,PTbook}.

In recent years, the interest in non-Hermitian physics has grown in the integrability community, mainly driven by recent developments in \emph{fishnet theory}. This theory, proposed in \cite{Gurdogan:2015csr}, is a deformation of $\mathcal{N}=4$ Super-Yang-Mills (SYM) whose Lagrangian is non-Hermitian. Similarly to $\mathcal{N}=4$ SYM, the action of the dilatation operator at one loop order on single-trace operators can be mapped to the action of a nearest-neighbour Hamiltonian on a chain of spins \cite{Ipsen:2018fmu}. This spin chain is called \emph{eclectic spin chain}, and its most interesting characteristic is that its Hamiltonian is non-diagonalisable, and thus, non-Hermitian. Although the eclectic spin chain is integrable, as its Hamiltonian can be obtained from an R-matrix, the usual Bethe Ansatz tools are not capable of retrieving the full Hilbert space \cite{StaudacherAhn}. Two different strategies to circumvent this problem have been posed: the first one uses the symmetries of the problem to compute the spectrum and Hilbert space combinatorially \cite{StaudacherAhn,Ahn:2021emp,Ahn:2022snr}, while the second one considers the non-diagonalisable Hamiltonian as the limit of a diagonalisable one and computes the limit of eigenvectors and eigenvalues in a specific way \cite{NietoGarcia:2021kgh,NietoGarcia:2022kqi}. Both approaches provide matching results.

The success in understanding the structure of the Hilbert space of the eclectic spin chain has motivated us to explore other non-diagonalisable integrable systems. Luckily for us, the interest in classifying all possible quantum R-matrices is nearly as old as the concept of the quantum inverse scattering method,\footnote{There exist different approaches to this problem: requiring the R-matrix to have certain symmetries \cite{Kulish:1981gi,Kulish1982,Jimbo:1985ua}, Baxterisation and representation theory \cite{Turaev:1988eb,Jones:1989ed,Jones:1990,Wu_1993}, transforming the Yang-Baxter equation into a differential equation \cite{Vieira:2017vnw,Vieira:2019vog}, reconstructing it from Hamiltonians that satisfy the Reshetikhin condition \cite{DeLeeuw:2019gxe,DeLeeuw:2019fdv,deLeeuw:2020ahe,deLeeuw:2020xrw,Corcoran:2023zax}, etc.} so we already have a list of candidates we can choose from. In this article, we will focus solely on one of the R-matrices found in \cite{DeLeeuw:2019gxe}, denoted as \emph{Class 5}. Although currently there are no known applications of this model, this does not mean that studying it is completely useless. We show that the spectrum of this model is relatively simple and can be understood in terms of the spectrum of the XXX spin chain. We believe that this simplicity makes the Class 5 model a good training ground where we can familiarise ourselves with the unusual properties of non-diagonalisable integrable models.

The outline of this article is as follows. In section 2 we review some concepts regarding non-diagonalisable matrices. In section 3, we present the Hamiltonian and the R-matrix of the Class 5 model. Sections 4 and 5 are dedicated to describe the structure of Jordan blocks of the transfer matrix and the Hamiltonian respectively. In section 6 we present some conclusions and future directions. We close our article with an appendix where we compute the behaviour of the transfer matrix for large values of the spectral parameter.

\section{Algebra of non-diagonalisable matrices}
\label{algebra}

We say that a matrix $M$ is \emph{non-diagonalisable} or \emph{defective} if the geometric multiplicity of one or more of its eigenvalues is smaller than their algebraic multiplicity. This means that we cannot construct as many eigenvectors as we expected from the multiplicity of roots in the characteristic polynomial. We cannot diagonalise this kind of matrices, so we have to resign ourselves to complete our basis of vectors in an intelligent way. This is done by means of generalised eigenvectors.

Given an eigenvector $v_{i,\alpha}^{(1)}$ associated with a defective eigenvalue $\lambda_i$, where the index $\alpha$ labels the possible geometric multiplicity of the eigenvalue $\lambda_i$, we define the \emph{generalised eigenvector of rank $n$ associated with $v_{i,\alpha}^{(1)}$} as the vector fulfilling
\begin{equation}
	(M-\lambda_i \mathbb{I} ) v_{i,\alpha}^{(n)}=v_{i,\alpha}^{(n-1)} \ . \label{geneigenvdefinition}
\end{equation}
This tower of generalised eigenvectors is called \emph{Jordan chain}. It is important to have in mind that these vectors are linearly independent, but not necessarily orthogonal. The number of vectors that form a Jordan chain is usually called the \emph{length} of said Jordan chain.

Thanks to these additional linearly independent vectors, we can write a similarity transformation $T$ that turns a given square matrix into a block diagonal matrix of the form
\begin{displaymath}
	T^{-1} M T=\begin{pmatrix}
	J_1 & 0 & 0 &   \dots \\
	0 & J_2 & 0 &  \dots \\
	0 & 0 & J_3  & \dots \\
	\vdots & \vdots & \vdots & \ddots
	\end{pmatrix} \ ,  \text{ with each block being of the form } J_i=\begin{pmatrix}
	\lambda_i & 1 & 0 &   \dots \\
	0 & \lambda_i & 1 &  \dots \\
	0 & 0 & \lambda_i & \dots \\
	\vdots & \vdots & \vdots  & \ddots
	\end{pmatrix} \ .
\end{displaymath}
This matrix is called the \emph{Jordan normal form} of $M$, and each block is called \emph{Jordan block} or \emph{Jordan cell}.\footnote{Jordan normal forms can also be constructed for matrices of generic dimension $N \times M$, however, we will only consider square matrices in this article.} The size of each Jordan block is equal to the number of generalised eigenvectors that form the Jordan chain associated with it.

However, we have to be careful with matrices that depend on a parameter, as the structure of Jordan blocks may change. As an example, consider
\begin{equation}
    M(\epsilon ) = \begin{pmatrix}
        3 & \epsilon & 1 \\
        0 & 3 & \epsilon \\
        0 & 0 & 3
    \end{pmatrix} \ .
\end{equation}
For generic values of $\epsilon$, we have a single Jordan chain of size 3
\begin{equation}
    v^{(1)}=(1,0,0) \ , \qquad v^{(2)}=\left( 0 , \frac{1}{\epsilon} , 0 \right) \ , \qquad v^{(3)}=\left( 0 , -\frac{1}{\epsilon^3} , \frac{1}{\epsilon^2} \right) \ .
\end{equation}
But, if we consider the point $\epsilon=0$, the Jordan chain splits into a Jordan chain of size 2 and a Jordan chain of size 1. We can see that, when properly normalised, $v^{(2)}$ and $v^{(3)}$ coalesce (i.e., become proportional) in that limit. It would be interesting to see if a procedure similar to the one described in \cite{NietoGarcia:2021kgh} can be applied to study this situation.

We should also stress that, similarly to what happens in diagonalisable matrices with degenerated eigenvalues, the linear combination of two generalised eigenvectors of rank $p$ associated with the same eigenvalue is also a generalised eigenvector of rank $p$, but the generalised eigenvector of rank $p-1$ it is associated with changes. On the other hand, the linear combination of a generalised eigenvector of rank $p$ and a generalised eigenvector of rank $q<p$ associated with the same eigenvalue is also a generalised eigenvector of rank $p$.

We want to end this section with some discussion about commuting matrices. It is well-known that, given two diagonalisable matrices that commute, $M$ and $N$, there exists a basis of vectors in which both matrices are diagonal. This happens because the action of $N$ on an eigenvector of $M$ has to give us an eigenvector of $M$ with the same eigenvalue as the original one
\begin{displaymath}
    M v_i =\lambda_i v_i \Longrightarrow M \, (N \, v_i)=N \, M \, v_i = \lambda_i (N v_i) \ .
\end{displaymath}
Thanks to this property, we can diagonalise $N$ in each eigenspace of $M$, giving us a basis that simultaneously diagonalises both matrices.

However, a similar property does not exist for defective matrices. In general, two non-diagonalisable matrices that commute cannot be simultaneously written in Jordan normal form. In fact, it is not guaranteed that a basis where one is written in Jordan normal form and the other is upper triangular actually exists. The best we can do is to find a basis where both matrices are upper triangular. As a consequence, finding a complete set of defective matrices that commute is not as powerful as in the diagonalisable case because these matrices may not have the same structure of Jordan blocks, and thus, of Jordan chains.

\section{The Class 5 Hamiltonian and R-matrix}

In this section, we present the Hamiltonian and the R-matrix of the Class 5 model described in \cite{DeLeeuw:2019gxe}.

The Class 5 model is a periodic spin chain with the following nearest-neighbour Hamiltonian\footnote{Compared to the definition in \cite{DeLeeuw:2019gxe}, we have subtracted a factor $a_1 \mathbb{I}_j \mathbb{I}_{j+1}$ from $h_{j,j+1}$. We have done so because we have also eliminated a global factor $1-a_1 u$ in the definition of the R-matrix.}
\begin{align}
    \mathcal{H} &=\sum_{j=1}^L h_{j,j+1} \ ,  \qquad \qquad h_{L,L+1}=h_{L,1} \ ,  \label{H}\\
    h_{j,j+1} &=2a_1 P_{j,j+1} +  \left( \frac{a_2-a_3}{2} \mathbb{I}_j+\frac{a_2+a_3}{2} \sigma^z_j \right) \sigma^+_{j+1} - \sigma^+_j \left( \frac{a_2-a_3}{2} \mathbb{I}_{j+1}+\frac{a_2+a_3}{2} \sigma^z_{j+1} \right)\notag \ , \\
    &=\begin{pmatrix}
        2 a_1 & a_2 & -a_2 & 0 \\
        0 & 0 & 2 a_1 & a_3 \\
        0 & 2 a_1 & 0 & -a_3 \\
        0 & 0 & 0 & 2 a_1
    \end{pmatrix} \notag 
\end{align}
where $L$ is the length of the chain, $\sigma^\alpha_j$ are the usual Pauli matrices acting on the site $j$ of the spin chain, and $P$ is the permutation operator. We can split the Hamiltonian into two different contributions, $h_{j,j+1}=h^{XXX}_{j,j+1}+h^{(5)}_{j,j+1}$, one equal to the well-known XXX spin chain Hamiltonian and a second contribution that vanishes when $a_2=a_3=0$. For clarity, in the remaining of this article we will refer to the first term as the \emph{undeformed} Hamiltonian and the second term as the \emph{deformation}.

Notice that the periodic boundary conditions allow us to add a local piece of the form $A_j - A_{j+1}$ to each of the nearest-neighbour terms of the local Hamiltonian, $h_{j,j+1}$, without altering the full Hamiltonian $\mathcal{H}$. This means that only one of the two deformation parameters is physical and we can set $a_2=a_3$. However, we will keep both parameters because we intend to use this property as a check for our computations.

It is easy to prove that this Hamiltonian is non-diagonalisable. This can be done by showing that $\mathcal{H}-\mathcal{H}^{XXX}$, that is, the deformation part, is nilpotent. This operator is obviously non-zero, but if we apply it $L+1$ times to any state, the pigeonhole principle tells us that every term will have at least two $\sigma^+$ operators acting on at least one site of the spin chain, making $(\mathcal{H}-\mathcal{H}^{XXX})^{L+1}=0$.

Apart from its non-diagonalisability, another important property of the Hamiltonian (\ref{H}) is that it is integrable, as it arises from the following R-matrix
\begin{align}
    R_{a,b}(u) &= \begin{pmatrix}
        1+2 a_1 u & a_2 u & -a_2 u & a_2 a_3 u^2 \\
        0 & 2 a_1 u & 1 & -a_3 u \\
        0 & 1 & 2 a_1 u & a_3 u \\
        0 & 0 & 0 & 1+2 a_1 u
    \end{pmatrix} \label{Rmatrix} \\
    &=2a_1 u\, \mathbb{I}_a \mathbb{I}_b+ P_{a,b} +  u \left( \frac{a_2+a_3}{2} \mathbb{I}_a+\frac{a_2-a_3}{2} \sigma^z_a \right) \sigma^+_b \notag \\
    &- u \,\sigma^+_a \left( \frac{a_2+a_3}{2} \mathbb{I}_b+\frac{a_2-a_3}{2} \sigma^z_b \right)+a_2 a_3 u^2 \sigma^+_a \sigma^+_b \ , \notag \\
    h_{a,b} &= \left. \left[ R_{a,b}(0) \right]^{-1} \frac{d R_{a,b}(u)}{du} \right|_{u\rightarrow 0} \ .
\end{align}
We have used the fact that this R-matrix is of difference form to write it in terms of one spectral parameter instead of two. Similarly to the Hamiltonian, we can split the R-matrix into the 6-vertex R-matrix and a second contribution that vanishes at $a_2=a_3=0$, $R_{a,b}=R_{a,b}^{XXX}+R_{a,b}^{(5)}$. Again, we will refer to the first piece as the undeformed part and to the second part as the deformation.

We can construct the monodromy and transfer matrices associated with this R-matrix in the usual way
\begin{align}
    T(u)&=\begin{pmatrix}
        A(u) & B(u) \\
        C(u) & D(u)
    \end{pmatrix}_0 = R_{0,1}(u) R_{0,2}(u) \dots R_{0,L-1}(u) R_{0,L}(u) \ ,  \\
    \tau (u)&=\text{tr}_0 \left[ T(u) \right]=A(u)+D (u) \ , \label{transfer}
\end{align}
where the trace $\text{tr}_0$ is taken only over the auxiliary space, labelled as $0$. Although we are able to construct these objects, it was shown in \cite{StaudacherAhn,NietoGarcia:2021kgh} that the usual Algebraic Bethe Ansatz procedure cannot access the full Hilbert space when applied to non-diagonalisable transfer matrices. In fact, it is not only incapable of finding any of the generalised eigenvectors, but it also cannot find all the eigenvectors of the transfer matrix. Nevertheless, this does not mean that the monodromy and transfer matrices are completely useless for us, as we will see in the next section.

\section{Jordan blocks of the class 5 transfer matrix}

In this section, we study the eigenvalues and generalised eigenvectors of the transfer matrix of the Class 5 model. We show that its eigenvalues are the same as the eigenvalues of the undeformed model, i.e. the XXX spin chain. We also show that its eigenvectors are the highest weight states of the $\mathfrak{su}(2)$ multiplets in which we can organise the eigenstates of the XXX spin chain. The relationship between generalised eigenvectors of rank higher than one and eigenstates of the XXX spin chain is more complex, but we can show that they are linear combinations of descendents. We will prove these statements in two steps. First, we prove them for the large $u$ limit of the transfer matrix. Then, we show that our results hold true also for finite values of $u$.

After carefully analysing the structure of the R-matrix (\ref{Rmatrix}), we find that the transfer matrix at large values of the spectral parameter can be expanded as
\begin{equation}
    \frac{\tau (u)}{(2 a_1 u)^L }= \mathbb{I} + \exp \left[ \frac{a_2 + a_3}{2 a_1} S^+ \right] + \frac{L}{2a_1 u} \mathbb{I} + \mathcal{O} (u^{-2})  \ , \label{tau}
\end{equation}
where $S^\alpha=\sum_{j=1}^L \sigma^\alpha_j$. We have collected the details of the proof in appendix~\ref{appendix} due to its length.

The strategy now is to take advantage of the fact that $S^+$ is the sum of raising operators in each site of the chain and write $\tau$ in an appropriate basis. For later convenience, we will pick as a basis the eigenvectors of the undeformed transfer matrix obtained from the Algebraic Bethe Ansatz, grouped by multiplets and ordered by their eigenvalue with respect to $S^z$ in decreasing order.

Although the Algebraic Bethe Ansatz is a well-known method (there exist many reviews on the topic, e.g., \cite{Faddeev:1996iy,Levkovich-Maslyuk:2016kfv}), it is worth presenting some details relevant for our proof. The central idea of the Algebraic Bethe Ansatz is that we can use the $B$ operator appearing on the monodromy matrix to construct all the eigenvectors of the transfer matrix. The XXX spin chain has a $\mathfrak{su}(2)$ symmetry generated by the global spin operators $S^\alpha$. As a consequence, the eigenvectors of the transfer matrix can be organised in $\mathfrak{su}(2)$ multiplets.\footnote{In fact, the Hilbert space $(\mathbb{C}^2)^L$ actually decomposes into the sum of tensor product of irreducible modules of $\mathfrak{su}(2)$ and $\mathcal{S}_L$, the symmetric group,\begin{displaymath}
    (\mathbb{C}^2)^L= \sum_{M=0}^{[L/2]} \mathfrak{V}_{[L-M,M]} \otimes \mathfrak{S}^{(L/2-M)} \ ,
\end{displaymath}
where $[L/2]$ is the integer part of $L/2$, $\mathfrak{S}^{(L/2-M)}$ is the $\mathfrak{su}(2)$ representation of spin $\frac{L}{2}-M$ and $\mathfrak{V}_{[L-M,M]}$ is the $\mathcal{S}_L$ representation associated to the Young diagram with $L-M$ boxes in the first row and $M$ in the second row. This property is called \emph{Schur-Weyl duality}. \label{SWd}} Each multiplet consists of a highest weight state, constructed by acting with $B$ operators with finite $u$ on a pseudo-vacuum, together with its descendants, obtained from the former by acting with $B$ operators with $u=\infty$. If the highest weight state is built using $M$ $B$ operators, the multiplet corresponds to an irreducible representation of $\mathfrak{su}(2)$ with spin $\frac{L}{2} -M$, and thus, it has dimension $L-2M+1$.

We have chosen the eigenvectors of the undeformed model as our basis because it allows us to map the problem of finding the Jordan block structure of (\ref{tau}) to a representation theory problem. In this basis, $S^+$ can only transform vectors of a given $\mathfrak{su}(2)$ multiplet into vectors of the same multiplet. Furthermore, $[S^z ,  S^+]= 2S^+$. Combining these two facts, it is clear that the transfer matrix at large values of $u$ is an upper triangular matrix in the basis we have chosen. Furthermore, it is the sum of a diagonal matrix whose entries are the eigenvalues of the undeformed transfer matrix, and a strictly upper triangular matrix that vanishes when we set $a_2=a_3=0$. Therefore, we can establish a one-to-one correspondence between the Jordan blocks of our transfer matrix and the multiplets of the undeformed model. This allows us to extract the number of Jordan blocks and their size from the number of multiples and their dimension. This is a well studied problem in representation theory, which we can borrow to claim that the Jordan normal form of our transfer matrix is made of $\binom{L}{M}-\binom{L}{M-1}$ Jordan blocks of size $L-2M+1$, for $0\leq M \leq \frac{L}{2}$ integer.\footnote{A quick way to prove this formula is as follows. In a spin chain with $L-M$ spins up and $M$ spins down, there are a total of $\binom{L}{M}$ linearly independent states. However, as long as $M\leq \frac{L}{2}$, $\binom{L}{M-1}$ of those states have to be $\mathfrak{su}(2)$ descendents of states with $M-1$ spins down, giving us $\binom{L}{M}-\binom{L}{M-1}$ highest weight states.}

We should stress that the generalised eigenvectors are not equal to the descendents, but to linear combinations of them. This is obvious from the fact that the vectors that form the multiplets are eigenvectors of $S^z$ but $[\tau(u) , S^z] \neq 0$. However, $S^+$ is of Jordan normal form in our basis, making the computation of the generalised eigenvectors of $\tau (u)$ rather simple.

Let us consider now what happens for generic values of $u$. We may worry that the structure we have just described gets spoiled by additional contributions from subleading orders in the expansion, but we will argue that this is not the case at any order.

First, we show that the transfer matrix is still the sum of a diagonal matrix and a strictly upper triangular matrix that vanishes when we set $a_2=a_3=0$. For that, we make use of an argument similar to the one used in appendix~\ref{appendix} for computing the large $u$ limit of the transfer matrix. To obtain the transfer matrix, we have to consider the part of the monodromy matrix that is proportional to the identity in the auxiliary space. This means that a term will contribute to the transfer matrix if and only if it has as many lowering operators as raising operators on the auxiliary space. Having a closer look at the R-matrix, we find that the only term that is proportional to $\sigma^-$ in the auxiliary space is a piece of the permutation operator
\begin{displaymath}
    P=\frac{\mathbb{I}\otimes \mathbb{I}+\sigma^x \otimes \sigma^x+\sigma^y \otimes \sigma^y+\sigma^z \otimes \sigma^z}{2}=\sigma^- \otimes \sigma^++\sigma^+ \otimes \sigma^-+\frac{\mathbb{I}\otimes \mathbb{I}+\sigma^z \otimes \sigma^z}{2} \ .
\end{displaymath}
Notice that the term we are interested in is of the form $\sigma^- \otimes \sigma^+$. Therefore, all the $\sigma^-$ operators appearing on the auxiliary space imply the presence of a $\sigma^+$ operator acting on one of the physical spaces. In contrast, the presence of a $\sigma^+$ on the auxiliary space implies the presence of a $\sigma^-$ operator acting on a physical space if and only if it does not contain the deformation parameters $a_2$ and $a_3$. This means that every term in the transfer matrix will either raise or maintain the $S^z$ eigenvalue of the states it acts on, making the transfer matrix upper triangular in the basis we proposed. Even more, we can see that the terms which contain at least one contribution from the deformation always raise the eigenvalue of $S^z$. These two facts combined show that the transfer matrix is still the sum of a diagonal matrix, equal to the undeformed transfer matrix, and a strictly upper triangular matrix that vanishes when we set $a_2=a_3=0$.

Regarding the generalised eigenvectors, it is easy to prove that the Jordan chains cannot mix multiplets with different total spin, defined as the eigenvalue of $S^2=\sum_{j=1}^L (\sigma^x_j)^2 + (\sigma^y_j)^2 + (\sigma^z_j)^2$. This happens because the transfer matrix commutes with the total spin operator, which is a consequence of the fact that the R-matrix is the sums of tensor products of Pauli matrices, so $[R_{0,j},S^2]=0$.

Mixing between vectors from different modules with the same eigenvalue with respect to $S^2$ can and will happen. We may worry that this can lead to changes in the length of the Jordan chains we computed, as it happened in the example presented in section~\ref{algebra}. However, we can show that this is not the case by adapting the argument used in \cite{Kato} to show that diagonalisable matrices that depend polynomially on a parameter can only have isolated exceptional points (i.e., they can only become non-diagonalisable for specific values of such parameter). Because the entries of the transfer matrix are polynomials on the spectral parameter (at least, for finite values of $L$), and the dimension of the kernel of $(\tau(u) -\lambda \mathbb{I})^n$ is an integer number for different integer values of $n$, these dimensions cannot be smooth functions of $u$. Therefore, they should remain equal to the values we have computed except for a finite number of values of $u$.

Similarly to the case of large values of $u$, the generalised eigenvectors that make the Jordan chains are not the vectors that form the multiplets of the XXX spin chain but a linear combination of them. The argument is exactly the same as before, but the relationship between these sets of vectors is more involved. Despite that, there are two notable cases where the relationship is immediate: the generalised eigenvectors of rank 1 and rank 2. It is trivial to prove that the eigenvectors will always coincide with the highest weight states of the irreducible representations because the transfer matrix can only raise or maintain the $S^z$ eigenvalue. Therefore, the eigenvector associated with the Jordan chain of length $L+1$ is the pseudo-vacuum of the Algebraic Bethe Ansatz, the state with all the spins up. Furthermore, the eigenvectors associated with each of the Jordan chains of length $L-1$ are the one-magnon eigenstates.\footnote{In this case, the discussion from the previous footnote is very clear. As our model is invariant under shifts by one lattice site, the one-magnon eigenstates are completely determined by their momentum. We have a spin chain with $L$ sites, which means that there are $L$ possible momenta. However, the state with vanishing momentum is a descendent of the pseudo-vacuum.} In the case of the generalised eigenvectors of rank 2, this is a consequence of the fact that the linear combination of a generalised eigenvector of rank 2 and its associated generalised eigenvector of rank 1 is still a generalised eigenvector of rank 2 with the same associated eigenvector.

\section{Jordan blocks of the class 5 Hamiltonian}

We may think that, by knowing the structure of the Jordan blocks of the transfer matrix, we immediately know the structure of the Hamiltonian, as the latter is the logarithmic derivative of the former, but that is not the case. As we discussed in section~\ref{algebra}, even if $\mathcal{H}$ and $\tau (u)$ commute, they do not have to have the same structure of Jordan chains. In fact, we also discussed in the same section that a matrix that depends on a parameter may change the structure of its Jordan blocks for some values of this parameter. For that reason, we have to separately analyse the Hamiltonian.

Let us start by showing that the transfer matrix and the Hamiltonian indeed have different Jordan block structure. The easiest way to do it is to prove that $(\mathcal{H}-\mathcal{H}^{XXX})^{L-1}=0$. This would mean that the largest Jordan block in the Hamiltonian can only be of size $L-1$, while we found that the transfer matrix has exactly one Jordan block of size $L+1$.

The form of the Hamiltonian and the fact that we cannot act with $\sigma^+$ twice on the same site of the spin chain imply that $(\mathcal{H}-\mathcal{H}^{XXX})^{L-1}$ can only give a non-trivial result if it acts on a state with either all or all-but-one spins down. Let us consider first the state with all spin downs, where we find that
\begin{align}
    &(\mathcal{H}-\mathcal{H}^{XXX}) | \downarrow \downarrow \dots \downarrow \rangle \propto \sum_{j=1}^L \left( \sigma_j^+ \sigma_{j+1}^z -\sigma_j^z \sigma_{j+1}^+ \right) | \downarrow \downarrow \dots \downarrow \rangle \notag \\ &=\sum_{j=1}^L \left( \sigma_j^+ \sigma_{j+1}^z -\sigma_{j-1}^z \sigma_{j}^+ \right) | \downarrow \downarrow \dots \downarrow \rangle=\sum_{j=1}^L \left( -\sigma_j^+ + \sigma_{j}^+ \right) | \downarrow \downarrow \dots \downarrow \rangle=0 \ . \label{firstargument}
\end{align}
This state was the highest rank state of the only Jordan chain of length $L+1$ of the transfer matrix, but it is an eigenstate of the Hamiltonian.

Let us check now what happens to states with one spin up. The calculation simplifies if we act on these states $L-2$ times with $(\mathcal{H}-\mathcal{H}^{XXX})$ before applying the same trick as before
\begin{align}
    &(\mathcal{H}-\mathcal{H}^{XXX}) \left[ (\mathcal{H}-\mathcal{H}^{XXX})^{L-2}  \sum_{j=1}^L \alpha_j \sigma^+_j | \downarrow \downarrow \dots \downarrow \rangle \right]=(\mathcal{H}-\mathcal{H}^{XXX}) \sum_{j=1}^L \beta_j \sigma^-_j | \uparrow \uparrow \dots \uparrow \rangle \notag \\
    &\propto \sum_{j=1}^L \sum_{k=1}^L \beta_j \left( \sigma_k^+ \sigma_{k+1}^z - \sigma_{k-1}^z \sigma_k^+ \right) \sigma^-_j | \uparrow \uparrow \dots \uparrow \rangle \notag \\
    &=\sum_{j=1}^L \beta_j \left[ (\sigma_j^+ \sigma_j^-) \sigma_{j+1}^z - \sigma_{j-1}^z (\sigma_j^+ \sigma_j^-) \right) | \uparrow \uparrow \dots \uparrow \rangle =0 \ .
\end{align}
As this result holds for any set of coefficients $\alpha_j$ (the relation between $\alpha$'s and $\beta$'s can be computed, but it is not important for us), the size of all Jordan blocks of the Hamiltonian is smaller than $L$.

This result strengthens in the case $\alpha_j=1$, where we get
\begin{align}
    &(\mathcal{H}-\mathcal{H}^{XXX})  \sum_{j=1}^L  \sigma^+_j | \downarrow \downarrow \dots \downarrow \rangle \propto \sum_{k=1}^L \left(\sigma^+_k \sigma^z_{k+1} - \sigma^z_k \sigma^+_{k+1} \right)\sum_{j=1}^L  \sigma^+_j | \downarrow \downarrow \dots \downarrow \rangle \ .
\end{align}
If $k$ is far from $j$, we can apply the same argument as in (\ref{firstargument}) and cancel those contributions. This leaves us only with the terms where $\sigma^+$ is applied at sites adjacent to the excitation, that is, $k=j-1$ or $k=j+1$. However, those cases also do not contribute, as we can cancel them thanks to the periodicity of the wavefunction
\begin{align*}
    \left[ (\sigma^+_{j-1} \sigma^z_{j}) \sigma^+_j + (-\sigma^z_{j-1} \sigma^+_{j}) \sigma^+_{j-1} \right]\downarrow \downarrow \dots \downarrow \rangle =0  \ , \\
    \left[ (-\sigma^z_{j-2} \sigma^+_{j-1}) \sigma^+_j + (\sigma^+_{j} \sigma^z_{j+1}) \sigma^+_{j-1} \right]\downarrow \downarrow \dots \downarrow \rangle =0 \ .
\end{align*}
A similar computation can be done \emph{mutatis mutandis} to any descendent of the pseudo-vacuum, i.e. any state of the form $\sum_{j_1 < j_2 < \dots < j_m} \sigma^+_{j_1} \sigma^+_{j_2} \dots \sigma^+_{j_m} | \downarrow \downarrow \dots \downarrow \rangle$, meaning that they are all eigenvectors of the Class 5 Hamiltonian. Consequently, the Jordan block of size $L+1$ that appears in the Jordan normal form of the transfer matrix, which we know is associated to descendents of the pseudo-vacuum, splits into $L+1$ Jordan blocks of size 1 in the Hamiltonian.

The mismatch between the transfer matrix and the Hamiltonian goes beyond the Jordan block associated to the pseudo-vacuum. Another example of this mismatch is the state $\sum_{j<k} \alpha(j,k) \sigma^-_j \sigma^-_k | \uparrow \uparrow \dots \uparrow \rangle$ with $\alpha (j,j+1)=0$. Not all two-magnon Bethe states can be written in that form, but a state of this form always exists for even values of $L$. This state is the first descendent of the one-magnon state with momentum $p=-\pi$, given by $\sum_{j} (-1)^{j} \sigma^-_j | \uparrow \uparrow \dots \uparrow \rangle$. We can indeed check that a state of this form is an eigenstate of $\mathcal{H}$, as
\begin{displaymath}
    (\mathcal{H}-\mathcal{H}^{XXX}) \sigma^-_j \sigma^-_k | \uparrow \uparrow \dots \uparrow \rangle \propto (\sigma^z_{j-1} \sigma^+_j - \sigma^+_j \sigma^z_{j+1} + j\rightarrow k) \sigma^-_j \sigma^-_k | \uparrow \uparrow \dots \uparrow \rangle=0 \ ,
\end{displaymath}
if $k\neq j \pm 1$. As a consequence, the Jordan chain associated with the one-magnon state with $p=-\pi$ splits into, at least, two Jordan chains.

We may think that this shortening of Jordan chains is due to some \emph{symmetry enhancement} that promotes some of the generalised eigenvectors to true eigenvectors. Although plausible, we can see that it cannot be due to the existence of normal subgroups of the symmetric group $\mathcal{S}_L$. In fact, the Schur-Weyl duality (see footnote~\ref{SWd}) of the undeformed chain tells us that the Hilbert space can be written as the sum of tensor products of irreducible representations of $\mathfrak{su}(2)$ and $\mathcal{S}_L$ modules. Thus, normal subgroups of $\mathcal{S}_L$ will never explain the split of Jordan blocks, which are in correspondence with irreducible representations of $\mathfrak{su}(2)$.

As a final comment, we want to point that the Jordan blocks of the Hamiltonian cannot be longer than the ones we found for the transfer matrix. This happens because the Jordan blocks of the transfer matrix are \emph{maximal} in the sense that the commutation with $S^+$ and $S^2$ forbids them from being longer. Their dimensions already coincide with the decomposition into irreducible representation of the tensor product of $L$ fundamental representations, and gluing two of them together would break one of the two symmetries.

\section{Conclusions}

In this article, we have untangled the structure of the Jordan chains that form the Hilbert space of the Class 5 model described in \cite{DeLeeuw:2019gxe}. By studying the asymptotic behaviour of the transfer matrix, we have found that the Jordan normal form of the transfer matrix is given by $\binom{L}{n}-\binom{L}{n-1}$ Jordan blocks of size $L+1-2n$, where $n$ is an integer number with $0\leq n \leq \frac{L}{2}$. We have also shown that the structure of Jordan blocks of the Hamiltonian is different from the one of the transfer matrix, being either of equal size or shorter. In particular, we have checked that the generalised eigenvectors that form the only Jordan chain of length $L+1$ of the transfer matrix are all eigenvectors of the Hamiltonian. This means that the Jordan block of size $L+1$ that appears in the transfer matrix splits into $L+1$ different Jordan blocks of size $1$ on the Hamiltonian. In striking difference with diagonalisable models, where commuting matrices can be simultaneously diagonalised, defective matrices that commute cannot be simultaneously brought to Jordan normal form. Consequently, having a set of defective operators that commute is not as powerful as if they were diagonalisable. In fact, the only information we get from our computation regarding the remaining conserved charges of the Class 5 model is that their Jordan chains cannot be longer than the ones of the transfer matrix.

There are several possible directions in which we can continue this project. The most obvious one would be to repeat our computations, but using instead the method detailed in \cite{NietoGarcia:2021kgh,NietoGarcia:2022kqi}. Here we approached the problem from a combinatorial perspective by using symmetries, so making sure that this other approach gives the same Hilbert space would provide a robust check of our results. However, said method requires the existence of a diagonalisable R-matrix with a tunable parameter that becomes the one we are interested in at a particular value of this parameter. Inspired by \cite{DeLeeuw:2019gxe,DeLeeuw:2019fdv,deLeeuw:2020ahe,deLeeuw:2020xrw}, we can instead construct a diagonalisable Hamiltonian with a tunable parameter that fulfils the Reshetikhin condition and use the Sutherland equations to find its corresponding R-matrix, as this seems like a simpler task.

Another possible direction would be to try to construct a new version of the Algebraic Bethe Ansatz that works for non-diagonalisable transfer matrices. The transfer matrix associated to the Class 5 model is simple enough that it could be a nice laboratory where we can test different ideas. In particular, the fact that all the eigenvectors of the Class 5 model are also eigenvectors of the undeformed model is specially useful, and it may help us shed some light on topics like the \emph{unexpected shortening} \cite{Ahn:2021emp} and \emph{misplaced eigenvectors} \cite{NietoGarcia:2021kgh}.

Finally, we have to remember that this is not the only non-diagonalisable R-matrix. In fact, the classifications \cite{DeLeeuw:2019gxe} and \cite{Corcoran:2023zax} offer us many choices. Thus, it would be interesting to study the structure of Jordan blocks of these other R-matrices.

\section*{Acknowledgments}

We thank R. Ruiz, L. Corcoran, and C. Paletta for useful discussions and for providing comments on a first version of this manuscript. In addition, we thank G. Arutyunov for useful discussions related to this work. JMNG is supported by the Deutsche Forschungsgemeinschaft (DFG, German Research Foundation) under Germany’s Excellence Strategy – EXC 2121 “Quantum Universe” – 390833306.

\appendix

\section{Large $u$ expansion of the transfer matrix} \label{appendix}

In this appendix, we prove that the transfer matrix of the Class 5 model is given by (\ref{tau}) for large values of the spectral parameter. We will provide the calculation of the leading order in full detail, but we will only sketch the computation for the next-to-leading order, as it is fairly similar.

For later convenience, let us start by writing the different terms of the R-matrix (\ref{Rmatrix}) as follows
\begin{equation}
    R_{0,a}(u) = R^{(0)}_{0,a}(u) + R^{(1,1)}_{0,a}(u) +R^{(1,2)}_{0,a}(u) +R^{(1,3)}_{0,a}(u) +R^{(2)}_{0,a}(u) \ ,
\end{equation}\vspace{-1cm}
\begin{align*}
   R^{(0)}_{0,a}(u) &=P_{0,a} \ , \qquad  R^{(1,1)}_{0,a}(u) =2a_1 u\, \mathbb{I}_0 \mathbb{I}_a \ , \\
   R^{(1,2)}_{0,a}(u) &=u \left( \frac{a_2+a_3}{2} \mathbb{I}_0+\frac{a_2-a_3}{2} \sigma^z_0 \right) \sigma^+_a \ , \\
   R^{(1,3)}_{0,a}(u) &=- u \,\sigma^+_0 \left( \frac{a_2+a_3}{2} \mathbb{I}_a+\frac{a_2-a_3}{2} \sigma^z_a \right) \ , \\
   R^{(2)}_{0,a}(u) &=a_2 a_3 u^2 \sigma^+_0 \sigma^+_a \ . \notag
\end{align*}
The transfer matrix is the product of $L$ R-matrices, each acting on the auxiliary space $0$ and a different physical site of the spin chain. To extract its leading order at large $u$ we start by identifying the leading order at large $u$ of the R-matrix, which corresponds to the term $R^{(2)}_{0,a}(u)=a_2 a_3 u^2 \sigma^+_a \sigma^+_b$. However, we cannot construct the transfer matrix only with terms of this form for two reasons: first because $(\sigma^+)^2=0$, and second because the terms proportional to $\sigma^+$ in the auxiliary space contribute to the operator $B$ but not to $A$ or $D$, and thus, not to the transfer matrix. Therefore, $R^{(2)}_{0,a}(u)$ can only appear in the product of R-matrices if another R-matrix contributes with a factor $\sigma^-_0$ to compensate the $\sigma^+_0$. The only piece of the R-matrix that contains $\sigma^-$ on the auxiliary space is $R^{(0)}_{0,a}(u) =P_{0,a}$, meaning that the product of R-matrices will only contribute to the transfer matrix if it has as many $R^{(2)}_{0,a}(u)$ as $R^{(0)}_{0,a}(u)$. However, this raises another problem. The product $R^{(0)}_{0,a}(u) R^{(2)}_{0,b}(u)$ is of order the same order as $R^{(1,x)}_{0,a}(u) R^{(1,y)}_{0,b}(u)$ for any $x$ and $y$. As a consequence, the leading order of the transfer matrix will be the sum over choosing between $R^{(0)}_{0,a}(u) R^{(2)}_{0,b}(u)$, $R^{(1,1)}_{0,a}(u)$ and $R^{(1,2)}_{0,b}(u)$ for each of the physical sites, i.e., schematically
\begin{equation}
    \tau (u)=\sum_{\substack{\text{all possible}\\ \text{ordering}}} \sum_{\substack{l,m,n \\ 2l+m+n=L}} \left( R^{(0)}_{0,a}(u) \right)^l \left( R^{(1,1)}_{0,a}(u) \right)^m \left( R^{(1,2)}_{0,a}(u) \right)^n \left( R^{(2)}_{0,a}(u) \right)^l \ ,
\end{equation}
where the first sum is taken over all possible ways of ordering the R-matrices such that each act on a different physical space. Notice that $R^{(1,3)}_{0,a}$ does not contribute at leading order because it acts as $\sigma^+$ on the auxiliary space, so it needs to be paired with a $R^{(0)}_{0,a}$ to contribute to the transfer matrix. As $R^{(0)}_{0,a} R^{(1,3)}_{0,a}$ is of order $u$ instead of $u^2$, it only contributes at next-to-leading order.

To simplify these sums, we can take advantage of the fact the number of $\sigma^+$ acting on a physical space that appear in a term is exactly the same as the number of deformation parameters, $a_2$ and $a_3$, it has.\footnote{One might argue that this is not true because $R^{(0)}_{0,a}$ contains terms with $\sigma^+$ but no deformation parameter. However, $R^{(0)}_{0,a}$ cannot appear at leading order without $R^{(2)}_{0,a}$, so this counting holds.} Thus, we can separate the different terms by how many powers of the deformation parameters they have. For example, it is clear that the term with no deformation parameters can only come from the product of $L$ terms of the form $R^{(1,1)}_{0,a}(u)=2a_1 u \mathbb{I}_0 \mathbb{I}_a$
\begin{equation}
    R^{(1,1)}_{0,1}(u) R^{(1,1)}_{0,2}(u) \dots R^{(1,1)}_{0,L}(u)= (2a_1 u)^{L} \, \mathbb{I}_0 \otimes \left( \bigotimes^L_{a=1} \mathbb{I}_a \right)= (2a_1 u)^{L} \, \mathbb{I}_0 \otimes \mathbb{I} \ . \label{basicproduct}
\end{equation}
This gives us a contribution of the form $(2a_1 u)^{L} \mathbb{I}$ to $A$ and another of the same form to $D$.

Terms linear in the deformation parameters can only come from replacing one of the factors $R^{(1,1)}_{0,a}$ in the term we just studied by $R^{(1,2)}_{0,a}$. In this case, the sum over all possible ordering of the R-matrices becomes a sum over which factor we replace
\begin{multline*}
    \sum_{a=1}^L \left[ (2a_1 u \mathbb{I}_0)^{a-1} u \left( \frac{a_2+a_3}{2} \mathbb{I}_0+\frac{a_2-a_3}{2} \sigma^z_0 \right) (2a_1 u \mathbb{I}_0)^{L-a} \right] \sigma^+_a=\\= (2a_1 u)^{L-1} u \left( \frac{a_2+a_3}{2} \mathbb{I}_0+\frac{a_2-a_3}{2} \sigma^z_0 \right) \sum_{a=1}^L \sigma^+_a \ .
\end{multline*}
The action on the physical space is nothing else but $S^+$, so this term contributes with $(2a_1 u)^{L-1} a_2 u S^+$ to the operator $A$ and with $(2a_1 u)^{L-1} a_3 u S^+$ to the operator $D$. Adding both, the total contribution to the transfer matrix is $(2a_1 u)^{L-1} (a_2 +a_3) u S^+$. As expected, it only depends on the sum of the two deformation parameters.

Similarly, the terms with two deformation parameters are constructed by replacing two of the factors $R^{(1,1)}_{0,a}$ in (\ref{basicproduct}) either by two factors of the form $R^{(1,2)}_{0,a} $ or by one factor of the form $R^{(0)}_{0,a} R^{(2)}_{0,b} + R^{(2)}_{0,a} R^{(0)}_{0,b}$. The first kind of substitution gives us a term of the form
\begin{multline*}
    \sum_{a=1}^L\sum_{b=a+1}^L \left[ (2a_1 u \mathbb{I}_0)^{a-1} u \left( \frac{a_2+a_3}{2} \mathbb{I}_0+\frac{a_2-a_3}{2} \sigma^z_0 \right) (2a_1 u \mathbb{I}_0)^{b-a-1} u \left( \frac{a_2+a_3}{2} \mathbb{I}_0+ \right.\right. \\
    \left. \left. \frac{a_2-a_3}{2} \sigma^z_0 \right) (2a_1 u \mathbb{I}_0)^{L-b} \right] \sigma^+_a \sigma^+_b= (2a_1 u)^{L-2} u^2 \left( \frac{a_2^2+a_3^2}{2} \mathbb{I}_0+\frac{a_2^2-a_3^2}{2} \sigma^z_0 \right) \sum_{a<b} \sigma^+_a \sigma^+_b \ ,
\end{multline*}
where the sum $\sum_{a<b}$ prevents us from overcounting terms. This term provides a contribution of the form $(2a_1 u)^{L-2} a_2^2 u^2 \sum_{a<b} \sigma^+_a \sigma^+_b$ to the operator $A$ and $(2a_1 u)^{L-2} a_3^2 u^2 \sum_{a<b} \sigma^+_a \sigma^+_b$ to the operator $D$. Furthermore, notice that $\sum_{a<b} \sigma^+_a \sigma^+_b$ can alternatively be written as $\frac{1}{2} (S^+)^2$. Regarding the second kind of substitutions, we can rewrite the auxiliary space part as
\begin{align*}
    P_{0,a} \sigma^+_0 &= \left(\sigma^-_0 \sigma^+_a + \sigma^+_0 \sigma^-_a + \frac{1}{2} \sigma^z_0 \sigma^z_a + \frac{1}{2} \mathbb{I}_0 \mathbb{I}_a \right) \sigma^+_0 =\frac{\mathbb{I}_0 - \sigma^z_0}{2} \sigma^+_a + \frac{1}{2} \sigma^z_0 \sigma^+_0  \sigma^z_a + \frac{1}{2} \sigma^+_0  \mathbb{I}_a \ .
\end{align*}
Of those terms, only the first one contributes to the transfer matrix. All the other terms have vanishing trace with respect to the auxiliary space. $\sigma^+_0 P_{0,a}$ can be treated similarly, giving us the exact same result. In total, after summing over all possible orderings, they contribute to the transfer matrix with a factor of the form $(2a_1 u)^{L-2} a_2 a_3 u^2 \sum_{a<b} \sigma^+_a \sigma^+_b$. Putting both kinds of terms together, the total contribution we get is $(2a_1 u)^{L-2} (a_2 + a_3)^2 u^2 \frac{(S^+)^2}{2}$.

Subsequent terms in the deformation parameters follow the same pattern. We have to substitute more and more factors $R^{(1,1)}_{0,a}$ in (\ref{basicproduct}) by one of the two options we have. At this point, the problem reduces to counting in how many ways we can order these different substitutions. There are $\binom{n+m}{n}$ different ways of ordering $n$ substitution of the first kind (i.e., one $R^{(1,1)}_{0,a} $ by one $R^{(1,2)}_{0,a} $) and $m$ of the second kind (i.e., two $R^{(1,1)}_{0,a} $ by $R^{(0)}_{0,a} R^{(2)}_{0,b} + R^{(2)}_{0,a} R^{(0)}_{0,b}$). Both substitutions give us the same contribution to the physical space, which amounts to
\begin{equation}
    \sum_{a_1<a_2<\dots < a_{n+2m}} \sigma^+_{a_1} \sigma^+_{a_2} \dots \sigma^+_{a_{n+2m}}=\frac{(S^+)^{n+2m}}{(n+2m)!} \ .
\end{equation}
After computing the product on the auxiliary space, we find that the sum over all orderings gives a contribution $\binom{n+m}{n} (2a_1 u)^{L-n-2m} a_2^n (a_2 a_3)^m u^{n+2m} \frac{(S^+)^{n+2m}}{(n+2m)!}$ to $A$ and a contribution with $a_2$ and $a_3$ swapped, $\binom{n+m}{n} (2a_1 u)^{L-n-2m} a_3^n (a_2 a_3)^m u^{n+2m} \frac{(S^+)^{n+2m}}{(n+2m)!}$, to $D$. We can now combine all the terms of the same order in the deformation parameter, which means that we have to sum over all contributions such that $n+2m=p$,
\begin{multline}
    \sum_{\substack{n,m\\ n+2m=p}} \binom{n+m}{n} (2a_1 u)^{L-n-2m} (a_2^n + a_3^n) (a_2 a_3)^m u^{n+2m} \frac{(S^+)^{n+2m}}{(n+2m)!}= \\ = (2a_1 u)^{L-p} (a_2+a_3)^p u^p \frac{(S^+)^{p}}{p!} \ .
\end{multline}

If we now put all the terms together, we obtain the following expression for the leading order at large $u$ of the transfer matrix
\begin{equation}
    \tau(u) = 2 (2a_1 u)^L + \sum_{p=1}^L (2a_1 u)^{L-p} (a_2+a_3)^p u^p \frac{(S^+)^{p}}{p!} + \mathcal{O} (u^{L-1}) \ .
\end{equation}
In addition, we can use the fact that our spin chain has $L$ sites, which implies $(S^+)^p=0$ for $p>L$, to extend the upper limit of the sum from $L$ to infinity. We can also absorb one of the factors $(2a_1 u)^L$ into the sum if we extend the lower limit of the sum to $p=0$. This allows us to formally write the sum as an exponential, giving us the leading order in (\ref{tau}). 

Now that we have constructed the leading order contribution, we can move to the next-to-leading order. The method to construct it is exactly the same as for the leading order, the only difference being that we have to substitute either one of the terms linear in $u$ by a term constant in $u$ or a term quadratic in $u$ by a term linear in $u$. This amounts to performing the following sums
\begin{align}
    \sum_{\substack{\text{all possible}\\ \text{ordering}}} \sum_{\substack{l,m,n \\ 2l+m+n=L}} &\left( R^{(0)}_{0,a}(u) \right)^l \left( R^{(1,1)}_{0,a}(u) \right)^{m+1} \left( R^{(1,2)}_{0,a}(u) \right)^n \left( R^{(2)}_{0,a}(u) \right)^{l-1} + \notag\\
    \sum_{\substack{\text{all possible}\\ \text{ordering}}} \sum_{\substack{l,m,n \\ 2l+m+n=L}} &\left( R^{(0)}_{0,a}(u) \right)^l \left( R^{(1,1)}_{0,a}(u) \right)^{m} \left( R^{(1,2)}_{0,a}(u) \right)^{n+1} \left( R^{(2)}_{0,a}(u) \right)^{l-1} + \notag\\
    \sum_{\substack{\text{all possible}\\ \text{ordering}}} \sum_{\substack{l,m,n \\ 2l+m+n=L}} &\left( R^{(0)}_{0,a}(u) \right)^l \left( R^{(1,1)}_{0,a}(u) \right)^{m} \left( R^{(1,2)}_{0,a}(u) \right)^n \left( R^{(1,3)}_{0,a}(u) \right) \left( R^{(2)}_{0,a}(u) \right)^{l-1} + \notag\\
    \sum_{\substack{\text{all possible}\\ \text{ordering}}} \sum_{\substack{l,m,n \\ 2l+m+n=L}} &\left( R^{(0)}_{0,a}(u) \right)^{l+1} \left( R^{(1,1)}_{0,a}(u) \right)^{m-1} \left( R^{(1,2)}_{0,a}(u) \right)^{n} \left( R^{(2)}_{0,a}(u) \right)^{l} + \notag\\
    \sum_{\substack{\text{all possible}\\ \text{ordering}}} \sum_{\substack{l,m,n \\ 2l+m+n=L}} &\left( R^{(0)}_{0,a}(u) \right)^{l+1} \left( R^{(1,1)}_{0,a}(u) \right)^{m} \left( R^{(1,2)}_{0,a}(u) \right)^{n-1} \left( R^{(2)}_{0,a}(u) \right)^{l} \ .
\end{align}
Despite its daunting look, we can apply the same logic we have used to compute the leading order. Instead of computing the sum as it is, we treat it as the combinatorial problem of counting all the possible ways we can replace the factors $R^{(1,1)}_{0,a}$ in (\ref{basicproduct}) by other contributions, considering terms with different powers of the deformation parameters separately.


Let us start from the term without deformation parameters. This term is constructed by replacing one of the factors $R^{(1,1)}_{0,a}$ in (\ref{basicproduct}) by $R^{(0)}_{0,a}$, and summing over all the possible replacements we can make. Because we want to compute the transfer matrix, only the piece $R^{(0)}_{0,a} \propto \frac{1}{2} \mathbb{I}_0 \mathbb{I}_a$ is relevant for this computation. The sum over all possible orderings is trivial to perform, giving us a factor of $L$. Thus, this term amounts to $\frac{L}{2} (2a_1 u)^{L-1} \mathbb{I}$ both for $A$ and $D$.

Let us consider now the case linear on the deformation parameters. In this case, we have to compute the following four possible products 
\begin{align*}
    &R^{(0)}_{0,a} R^{(1,2)}_{0,b}+R^{(0)}_{0,a} R^{(1,3)}_{0,b}+R^{(1,2)}_{0,a} R^{(0)}_{0,b} +R^{(1,3)}_{0,a} R^{(0)}_{0,b} =\\
    &P_{0,a} \left[ u \left( \frac{a_2+a_3}{2} \mathbb{I}_0+\frac{a_2-a_3}{2} \sigma^z_0 \right) \sigma^+_b - u \,\sigma^+_0 \left( \frac{a_2+a_3}{2} \mathbb{I}_b+\frac{a_2-a_3}{2} \sigma^z_b \right) \right] \\
    &+ \left[ u \left( \frac{a_2+a_3}{2} \mathbb{I}_0+\frac{a_2-a_3}{2} \sigma^z_0 \right) \sigma^+_a - u \,\sigma^+_0 \left( \frac{a_2+a_3}{2} \mathbb{I}_a+\frac{a_2-a_3}{2} \sigma^z_a \right) \right]P_{0,b} \ .
\end{align*}
After some algebra, we see that the final result is proportional to $\sigma^z$ on the auxiliary space, so it does not contribute to the transfer matrix.

For the quadratic order in the deformation parameters, we have the same two contributions we analysed at leading order, but substituting one of the $R^{(1,1)}_{0,a}$ by $R^{(0)}_{0,a}$, plus a contribution where we substitute $R^{(2)}_{0,a}$ by $R^{(1,3)}_{0,a}$. After some algebra, we see that the total contribution also vanishes. Higher orders in the deformation parameter follow a similar pattern of cancellations.

Thus, at next-to-leading order, we only get a piece of the form $L (2a_1 u)^{L-1}$. In total, the transfer matrix at order $u^{L-1}$ is
\begin{equation}
    \tau(u) = (2a_1 u)^L \mathbb{I} + (2a_1 u)^L \exp \left[ \frac{a_2 + a_3}{2 a_1} S^+ \right] + L (2a_1 u)^{L-1} \mathbb{I} + \dots \ .
\end{equation}



\begin{thebibliography}{99}

\bibitem{Ashida:2020dkc}
Y.~Ashida, Z.~Gong and M.~Ueda, 
``Non-Hermitian physics,'' 
Adv. Phys. \textbf{69} (2021) no.3, 249-435
[arXiv:2006.01837 [cond-mat.mes-hall]].


\bibitem{Bender:1998ke}
C.~M.~Bender and S.~Boettcher, 
``Real spectra in Non-Hermitian Hamiltonians having PT symmetry,'' 
Phys. Rev. Lett. \textbf{80} (1998), 5243-5246
[arXiv:physics/9712001 [physics]].

\bibitem{Bender:2007nj}
C.~M.~Bender, 
``Making sense of non-Hermitian Hamiltonians,'' 
Rept. Prog. Phys. \textbf{70} (2007), 947
[arXiv:hep-th/0703096 [hep-th]].



\bibitem{Alexandre:2015kra}
J.~Alexandre, C.~M.~Bender and P.~Millington, 
``Non-Hermitian extension of gauge theories and implications for neutrino physics,'' 
JHEP \textbf{11} (2015), 111
[arXiv:1509.01203 [hep-th]].


\bibitem{PTbook}
C.~M.~Bender, with contributions from P.~E.~Dorey, C.~Dunning, A.~Fring, D.~W.~Hook, H.~F.~Jones, S.~Kuzhel, G.~Lévai, and R.~Tateo, 
``PT Symmetry in Quantum and Classical Physics,'' 
World Scientific, 2018.


\bibitem{Gurdogan:2015csr}
\"O.~G\"urdo\u{g}an and V.~Kazakov, 
``New Integrable 4D Quantum Field Theories from Strongly Deformed Planar $\mathcal N = $ 4 Supersymmetric Yang-Mills Theory,'' 
Phys. Rev. Lett. \textbf{117} (2016) no.20, 201602
[arXiv:1512.06704 [hep-th]].


\bibitem{Ipsen:2018fmu}
A.~C.~Ipsen, M.~Staudacher and L.~Zippelius,
``The one-loop spectral problem of strongly twisted $ \mathcal{N} $ = 4 Super Yang-Mills theory,''
JHEP \textbf{04} (2019), 044
[arXiv:1812.08794 [hep-th]].

\bibitem{StaudacherAhn}
C.~Ahn and M.~Staudacher,
``The Integrable (Hyper)eclectic Spin Chain,''
JHEP \textbf{02} (2021), 019
[arXiv:2010.14515 [hep-th]].


\bibitem{Ahn:2021emp}
C.~Ahn, L.~Corcoran and M.~Staudacher,
``Combinatorial solution of the eclectic spin chain,''
JHEP \textbf{03} (2022), 028
[arXiv:2112.04506 [hep-th]].

\bibitem{Ahn:2022snr}
C.~Ahn and M.~Staudacher,
``Spectrum of the hypereclectic spin chain and P\'olya counting,''
Phys. Lett. B \textbf{835} (2022), 137533
[arXiv:2207.02885 [hep-th]].


\bibitem{NietoGarcia:2021kgh}
J.~M.~Nieto Garc\'\i{}a and L.~Wyss,
``Jordan blocks and the Bethe Ansatz I: The eclectic spin chain as a limit,''
Nucl. Phys. B \textbf{981} (2022), 115860
[arXiv:2112.13883 [hep-th]].

\bibitem{NietoGarcia:2022kqi}
J.~M.~Nieto Garc\'\i{}a,
``Jordan blocks and the Bethe Ansatz II: The eclectic spin chain beyond K = 1,''
JHEP \textbf{12} (2022), 106
[arXiv:2206.08348 [hep-th]].


\bibitem{Kulish:1981gi}
P.~P.~Kulish, N.~Reshetikhin and E.~Sklyanin,
``Yang-Baxter Equation and Representation Theory. 1.'',
Lett.~Math.~Phys. \textbf{5}, 393 (1981).

\bibitem{Kulish1982}
P.~P.~Kulish and E.~K.~Sklyanin,
``Solutions of the Yang-Baxter equation'',
Journal~of~Soviet~Mathematics \textbf{19}, 1596 (1982).

\bibitem{Jimbo:1985ua}
M.~Jimbo,
``Quantum r Matrix for the Generalized Toda System'',
Commun.~Math.~Phys. \textbf{102}, 537 (1986).



\bibitem{Turaev:1988eb}
V.~Turaev,
``The Yang-Baxter equation and invariants of links'',
Invent.~Math. \textbf{92}, 527 (1988).

\bibitem{Jones:1989ed}
V.~Jones,
``On knot invariants related to some statistical mechanical models'',
Pacific~J.~Math. \textbf{137}, 311 (1989).

\bibitem{Jones:1990}
V.~Jones,
``Baxterization'',
International~Journal~of~Modern~Physics~B \textbf{4}, 701 (1990).

\bibitem{Wu_1993}
F.~Wu,
``The Yang-Baxter Equation in Knot Theory'',
International~Journal~of~Modern~Physics~B  \textbf{07}, 3737 (1993).

\bibitem{Vieira:2017vnw}
R.~S.~Vieira,
`Solving and classifying the solutions of the Yang-Baxter equation through a differential approach. Two-state systems,''
JHEP \textbf{10} (2018), 110
[arXiv:1712.02341 [nlin.SI]].

\bibitem{Vieira:2019vog}
R.~S.~Vieira,
``Fifteen-vertex models with non-symmetric $R$ matrices,''
[arXiv:1908.06932 [nlin.SI]].


\bibitem{DeLeeuw:2019gxe}
M.~De Leeuw, A.~Pribytok and P.~Ryan,
``Classifying two-dimensional integrable spin chains,''
J. Phys. A \textbf{52} (2019) no.50, 505201
[arXiv:1904.12005 [math-ph]].


\bibitem{DeLeeuw:2019fdv}
M.~De Leeuw, A.~Pribytok, A.~L.~Retore and P.~Ryan,
``New integrable 1D models of superconductivity,''
J. Phys. A \textbf{53} (2020) no.38, 385201
[arXiv:1911.01439 [math-ph]].

\bibitem{deLeeuw:2020ahe}
M.~de Leeuw, C.~Paletta, A.~Pribytok, A.~L.~Retore and P.~Ryan,
``Classifying Nearest-Neighbor Interactions and Deformations of AdS,''
Phys. Rev. Lett. \textbf{125} (2020) no.3, 031604
[arXiv:2003.04332 [hep-th]].

\bibitem{deLeeuw:2020xrw}
M.~de Leeuw, C.~Paletta, A.~Pribytok, A.~L.~Retore and P.~Ryan,
``Yang-Baxter and the Boost: splitting the difference,''
SciPost Phys. \textbf{11} (2021), 069
[arXiv:2010.11231 [math-ph]].


\bibitem{Corcoran:2023zax}
L.~Corcoran and M.~de Leeuw,
``All regular $4 \times 4$ solutions of the Yang-Baxter equation,''
[arXiv:2306.10423 [hep-th]].

\bibitem{Faddeev:1996iy}
L.~D.~Faddeev,
`How algebraic Bethe ansatz works for integrable model,''
[arXiv:hep-th/9605187 [hep-th]].

\bibitem{Levkovich-Maslyuk:2016kfv}
F.~Levkovich-Maslyuk,
``The Bethe ansatz,''
J. Phys. A \textbf{49}, no.32, 323004 (2016)
[arXiv:1606.02950 [hep-th]].

\bibitem{Kato}
T.~Kato, ``Perturbation Theory for Linear Operators,'' Springer-Verlag, 1966.

\end{thebibliography}
\end{document}